\documentclass{ifacconf}

\usepackage{graphicx}      
\usepackage{natbib}        
\usepackage{amsfonts,amsmath}
\usepackage[noend]{algpseudocode}
\usepackage{algorithm}

\algrenewcommand\algorithmicrequire{\textbf{Input:}}
\algrenewcommand\algorithmicensure{\textbf{Output}:}
\algrenewcommand\algorithmicindent{1.25em}

\begin{document}
\begin{frontmatter}

\title{
	On-Line Synthesis of Permissive Supervisors for Partially Observed Discrete Event Systems under scLTL Constraints\thanksref{footnoteinfo}
} 

\thanks[footnoteinfo]{
	This work was supported by JST ERATO Grant Number JPMJER1603, Japan, and JSPS KAKENHI Grant Number JP19J13487, Japan. 
}

\author[OU,DC]{Ami Sakakibara} 
\author[OU]{Toshimitsu Ushio} 

\address[OU]{
	Graduate School of Engineering Science, Osaka University, 
	1-3 Machikaneyama, Toyonaka, Osaka, 560-8531, Japan 
	(e-mail: sakakibara@hopf.sys.es.osaka-u.ac.jp; ushio@sys.es.osaka-u.ac.jp)
}
\address[DC]{JSPS Research Fellow}

\begin{abstract}                
We consider a supervisory control problem of a discrete event system (DES) under partial observation, where a control specification is given by a fragment of linear temporal logic. 
We design an on-line supervisor that dynamically computes its control action with the complete information of the product automaton of the DES and an acceptor for the specification. 
The concepts of controllability and observability are defined by means of a ranking function defined on the product automaton, which decreases its value if an accepting state of the product automaton is being approached.
The proposed on-line control scheme leverages the ranking function and a permissiveness function, which represents a time-varying permissiveness level. 
As a result, the on-line supervisor achieves the specification, being aware of the tradeoff between its permissiveness and acceptance of the specification, if the product automaton is controllable and observable. 
\end{abstract}


\end{frontmatter}

\section{Introduction}

The supervisory control theory for discrete event systems (DESs) has been widely studied since its initiation in \cite{Ramadge1987}. 
Synthesis of supervisors turns out to be computationally hard when the controlled system is large or has much complex aspects.
Researchers have overcome the difficulty by designing supervisors on-line. 
\cite{Chung1992,Chung1993,Chung1994} proposed a method to generate limited lookahead trees on-the-fly instead of constructing a complete supervisor and their methods are extended to the settings of partial observation \citep{Hadj-Alouane1996} or time-varying DESs \citep{Grigorov2006}.
Another way is taken to design on-line supervisors for partially observed DESs, where the supervisor modifies appropriate control actions precomputed in the case of full observation \citep{Heymann1994,Prosser1998}.

In the supervisory control framework, control requirements are typically given by formal languages, i.e., subsets of event sequences generated by the system.  
Practically, we need to translate desired behavior of the system into formal languages, which is a hard task.
For this reason, linear temporal logic (LTL) is paid much attention to as a formal specification language for control problems, thanks to its rich expressiveness \citep{Belta2017,Tumova2016,Jiang2006}. 
It is practically acceptable to restrict the specification language to a fragment of LTL like syntactically co-safe LTL (scLTL), for which synthesis problems can be solved in much less complexity than for the case of the general LTL \citep{Kupferman2001}.

In \cite{Sakakibara2019}, we consider a supervisory control problem of a DES under an scLTL constraint.
We propose an on-line control scheme, where we leverage a \textit{ranking function} that enables us to find desirable behavior with respect to the scLTL specification. 
The concept of ranking functions is like that of Lyapunov functions, which play a great role in determining control strategies.
The key idea is that, if the rank decreases along a trajectory, we regard it as good.
Ranking functions are useful in solving games played on a graph and reachability analysis of automata, which sometimes give solutions to LTL-related problems. 
We define a ranking function on the product automaton of the DES and the specification automaton so that its value decreases 
if an accepting state of the product automaton is being approached. 
To take the tradeoff between permissiveness of the supervisor and acceptance of the specification, 
we additionally introduce a \textit{permissiveness function} that indicates a time-varying permissiveness level. 
If we have a higher permissive level, the supervisor may enable events that do not necessarily lead to  achievement of the specification. 
By referring to the permissiveness level together with the ranking function at each step of the on-line control process, the supervisor computes more permissive control patterns. 

This paper extends the on-line supervisory control scheme proposed in \cite{Sakakibara2019} to the setting of partial observation. 
After each observation, the supervisor dynamically computes its control action with the information of the fully observed product automaton, on which the ranking function is defined. 
Furthermore, we characterize the concepts of controllability and observability by means of the ranking function.
The supervisor forces the DES to satisfy the scLTL specification if the product automaton is controllable and observable.

The rest of this paper is organized as follows. 
Section \ref{sec:preliminaries} gives fundamental definitions and notations. 
Then, Section \ref{sec:formulation} formulates a supervisory control problem for scLTL specifications. 
Section \ref{sec:ranking} briefly explains the ranking function with its related properties.  
Section \ref{sec:on-line} proposes our on-line control scheme, which is demonstrated in Section \ref{sec:example} with a simple example.
Finally, Section \ref{sec:conclusion} concludes the paper. 

\section{PRELIMINARIES}

\label{sec:preliminaries}

For a set $T$, we denote by $ |T| $ its cardinality. 
$T^*$ (resp., $T^\omega$) represents a set of finite (resp., infinite) sequences over $T$. 
For a finite or infinite sequence $\tau$ over $T$, 
let $\tau[j]$ be the $(j+1)$-st element of $\tau$.
For any $ j,k \in \mathbb{N} $ with $ j \leq k $, we denote by $ \tau[j \ldots k] $ the sequence $ \tau[j]\tau[j+1]\ldots\tau[k] $. 
We write $ \tau' \preceq \tau $ if $ \tau' $ is a prefix of $ \tau $. 
For a finite sequence $ \tau \in T^* $, $ \|\tau\| $ stands for the length of $ \tau $.

\subsection{Discrete Event Systems}
A discrete event system (DES) is a tuple 
\[
G=((X,\Sigma,\delta,x_0), AP, L), 
\] 
where 
$X$ is the set of states, 
$\Sigma$ is the set of events, 
a partial function $\delta: X \times \Sigma \to X$ is the transition function, 
$x_0 \in X$ is the initial state, 
$AP$ is the set of atomic propositions, and 
$L: X \to 2^{AP}$ is the labeling function.
$G$ is said to be finite if $X$, $\Sigma$, and $AP$ are all finite. 
We write $\delta(x,\sigma)!$ if a transition from $x$ with $\sigma$ is defined. 
For each $x \in X$, let $\Sigma(x)=\{ \sigma \in \Sigma: \delta(x, \sigma)! \}$.
We denote by a triple $ (x,\sigma,x') $ a transition from $ x \in X $ to $ x' = \delta(x,\sigma) $ for some $ \sigma \in \Sigma(x) $. 
Moreover, the transition function is extended to a sequence of inputs: for $\sigma \in \Sigma$ and $s \in \Sigma^*$, 
$ \delta(x,\varepsilon) = x $ and $\delta(x,s\sigma)= \delta( \delta(x,s),\sigma)$.

Let $ \mathcal{L}(G) = \{ s \in \Sigma^*: \delta(x_0,s)! \} $ be the set of all finite event sequences generated by $ G $. 
An infinite sequence $ \rho \in X(\Sigma X)^\omega $ is called a {\it run} if, for any $j \in \mathbb{N}$, $\rho[2(j+1)] \in \delta(\rho[2j], \rho[2j+1])$.
A finite sequence $ h \in X(\Sigma X)^* $ is called a {\it history} if $ h \in X $ or, for $ h $ with $ \|h\|\geq 3 $, $  h[2(j+1)] \in \delta (h[2j], h[2j+1]) $ for any $j \in \{ 0, 1, \ldots, \frac{\|h\|-3}{2} \}$. 
The set of runs (resp., histories) starting from the initial state $ x_0 $ is defined as $ \mathsf{Runs}(G) $ (resp., $ \mathsf{His}(G) $).
DES $ G $ is said to be {\it deadlock-free} if, for any $ s \in \mathcal{L}(G) $, there exists $ \sigma \in \Sigma $ such that $ \delta(x_0,s\sigma)! $. 

The event set is partitioned into disjoint subsets $\Sigma = \Sigma_c \cup \Sigma_{uc}$, where $\Sigma_c$ (resp., $\Sigma_{uc}$) is the set of controllable (resp., uncontrollable) events.
We define $ \Sigma_c(x) = \Sigma(x) \cap \Sigma_c $ and $ \Sigma_{uc}(x) = \Sigma(x) \cap \Sigma_{uc} $ for each $ x \in X $. 
We have another partition of the event set $ \Sigma = \Sigma_o \cup \Sigma_{uo} $ with the set $ \Sigma_o $ (resp., $ \Sigma_{uo} $) of observable (resp., unobservable) events.
Let $ \mathcal{P}: \Sigma^* \to \Sigma_o^* $ be a natural projection defined inductively as follows:
\begin{align*}
&\mathcal{P}(\varepsilon) = \varepsilon, \\
&\forall s \in \Sigma^*, \forall \sigma \in \Sigma, \ \mathcal{P}(s\sigma ) = 
\begin{cases}
\mathcal{P}(s) \sigma & \text{ if } \sigma \in \Sigma_o, \\
\mathcal{P}(s)  & \text{ if } \sigma \in \Sigma_{uo}. \\
\end{cases}
\end{align*}
We also define the inverse $ \mathcal{P}^{-1}: 2^{\Sigma_o^*} \to 2^{\Sigma^*} $ as $ \mathcal{P}^{-1} (T) = \{ s \in \Sigma^*: t \in T, \mathcal{P}(s) = t \} $. 
The observable behavior of the DES $ G $ is given by $ \mathcal{P}(\mathcal{L}(G)) = \{ s = \mathcal{P}(t) \in \Sigma_o^*: t \in \mathcal{L}(G) \} $.



For the set $AP$, a {\it letter} $\nu \in 2^{AP}$ stands for a subset of atomic propositions.
A {\it word} is a finite or infinite string of letters. 
Each run in $\mathsf{Runs}(G)$ generates a sequence of letters, namely a word over $2^{AP}$, obtained by the labeling function.
We extend the labeling function to express such words: for run $\rho$,
$L(\rho)= L(\rho[0]) L(\rho[2]) \ldots $.
The extension for histories is defined similarly.

\subsection{Syntactically Co-Safe Linear Temporal Logic}

Linear temporal logic (LTL) is useful to describe qualitative control specifications. 
In this paper, we focus on syntactically co-safe LTL (scLTL), a subclass of LTL.
Formally, an scLTL formula $\varphi$ over the set $ AP $ of atomic propositions is defined as
\[
\varphi \! ::= \! \mathrm{true} \ | \  a \ |\ \neg a \ |\ \varphi_1 \land \varphi_2 \ | \ \varphi_1 \lor \varphi_2 \ | \ \bigcirc \varphi \ | \ \varphi_1 \! \mathbf{U} \varphi_2,  
\] 
where $ a \in AP $, $\varphi, \varphi_1, \varphi_2$ are scLTL formulas. 
In addition, we usually use a temporal operator $\Diamond$, which is defined by 
$ \Diamond\varphi := \mathrm{true} \mathbf{U} \varphi $.

The semantics of scLTL is defined over an infinite word \citep{Baier2008}.
For an scLTL formula $ \varphi $ over $ AP $ and an infinite word $ w \in (2^{AP})^\omega $, we write $ w \models \varphi $ if $ w $ satisfies $ \varphi $. 
For a DES $ G $ and an scLTL formula $ \varphi $, we say $ G $ satisfies $ \varphi $, denoted by $ G \models \varphi $, if $ L(\rho) \models \varphi $ for all $ \rho \in \mathsf{Runs}(G) $. 

Although scLTL formulas are evaluated over infinite words, it is known that we only need to check whether an input word has a \textit{good prefix} of the formula. 
Any scLTL formula can be translated into a corresponding deterministic finite automaton (DFA), which is an acceptor for the good prefixes \citep{Kupferman2001}.
For an scLTL formula $ \varphi $, let $ A_\varphi = ((X_A, \Sigma_A, \delta_A, x_{A,0}), F_A) $ be its corresponding DFA, where 
$ X_A $ is the set of states, 
$ \Sigma_A = 2^{AP} $ is the input alphabet, 
a total function $ \delta_A: X_A \times \Sigma_A \to X_A $ is the transition function, 
$ x_{A,0} \in X_A $ is the initial state, and 
$ F_A \subseteq X_A $ is the set of accepting states. 
For any $ w \in (2^{AP})^\omega $, we have 
\begin{equation}
\label{eq:acc_scLTL}
w \models \varphi \iff \exists w' \in (2^{AP})^*, w' \preceq w \land \delta_A(x_{A,0},w') \in F_A.
\end{equation}

\section{FORMULATION}
\label{sec:formulation}
In this paper, we formulate a supervisory control problem with scLTL specifications. 
A controller, called a \textit{supervisor}, enables some controllable events at each state \citep{Ramadge1987}. 
For each state $ x \in X $, we define $ \Gamma(x) = \{ \gamma \subseteq \Sigma(x): \Sigma_{uc} (x) \subseteq \gamma \} $, 
where 
$ \gamma \in \Gamma (x) $ is called a \textit{control pattern} at $ x $. 
Let $ \Gamma=\bigcup_{x \in X} \Gamma(x)$ be the set of all control patterns. 
The supervisor determines a control pattern after each observation. 
Formally, we define a \textit{supervisor under partial observation} $ \mathcal{P} $ as a mapping $ \mathcal{S}: \mathcal{P} (\mathcal{L}(G)) \to \Gamma $.

\begin{defn}[Supervised behavior]
	Let $ \mathcal{S} $ be a supervisor under partial observation $ \mathcal{P} $ for a DES $ G = ((X, \Sigma, \delta, x_0), AP, L) $. 
	The closed-loop behavior of the DES $ G $ under the control by $ \mathcal{S} $, denoted by $ \mathcal{S}/G $, is given by 
	\begin{align*}
	&\varepsilon \in \mathcal{L}(\mathcal{S}/G), \text{ and }\\
	&\forall s \in \mathcal{L}(\mathcal{S}/G), s\sigma \in \mathcal{L}(\mathcal{S}/G)\\
	&\qquad \qquad \qquad \qquad \iff s \sigma \in \mathcal{L}(G)  \land \sigma \in \mathcal{S}(\mathcal{P}(s)).  \\
	&\mathsf{Runs}(\mathcal{S}/G) = \{ x_0 \sigma_1 x_1 \ldots \in \mathsf{Runs}(G): \sigma_1 \in \mathcal{S}(\varepsilon) \\ 
	&\qquad \qquad \qquad \qquad \land \forall j \geq 1, \sigma_{j+1} \in \mathcal{S}(\mathcal{P}(\sigma_1 \sigma_2 \ldots \sigma_{j})) \}. 
	\end{align*}
\end{defn}
%
%

\begin{prob}
	\label{prob:main}
	Given a finite deadlock-free DES $ G=((X,\Sigma,\delta,$ $x_0), AP, L) $ and an scLTL formula $ \varphi $ over $ AP $, 
	synthesize a supervisor $ \mathcal{S} $ under partial observation $ \mathcal{P} $ such that $ \mathcal{S}/G \models \varphi $.
\end{prob}

To solve Problem \ref{prob:main}, we design an on-line supervisor, which dynamically computes a control pattern after observing an event occurrence. 
Our control scheme is divided into two stages; we first execute the preprocessing off-line, and then move on to the on-line control stage, where it stops controlling the DES after detecting a history corresponding to a good prefix of the scLTL specification.

\section{RANKING FUNCTION}
\label{sec:ranking}

The specification given by an scLTL formula $ \varphi $ is translated into the equivalent DFA $ A_\varphi $. 
Then, we obtain the product automaton $ P $ of the DES $ G $ and the DFA $ A_\varphi $, which is computed as follows. 
\[
P = G \otimes A_\varphi = ((X_P,\Sigma_P,\delta_P,x_{P,0}),F_P),
\]
where 
$ X_P = X \times X_A $ is the set of states, 
$ \Sigma_P = \Sigma $ is the set of events, 
$ \delta_P:X_P \times \Sigma_P \to X_P $ is the transition function, 
$ x_{P,0} = (x_0,\delta_A(x_{A,0},L(x_0))) $ is the initial state, and 
$ F_P = X \times F_A $ is the set of accepting states. 
For each $ x = (x_G,x_A) \in X_P $ and $ \sigma \in \Sigma_P $, 
$ \delta_P(x,\sigma) = \big(\delta(x_G,\sigma), \delta_A(x_A, L(\delta(x_G,\sigma)))\big) $ and we define $ J_G(x) = x_G $ and $ J_A(x) = x_A $.
The two kinds of event partitions are inherited from the DES: $ \Sigma_{P,c}=\Sigma_c $, $ \Sigma_{P,uc} = \Sigma_{uc} $, $ \Sigma_{P,o} = \Sigma_o $, and $ \Sigma_{P,uo} = \Sigma_{uo} $.
Note that $ \mathcal{L}(P) = \mathcal{L}(G) $ and $ \mathcal{P}(\mathcal{L}(P)) = \mathcal{P}(\mathcal{L}(G)) $. 

Since the product automaton captures the behavior of the DES and the DFA at the same time, our goal turns out to reach an accepting state of the product automaton. 
For that purpose, we introduce a ranking function with the existence of uncontrollable transitions \citep{Sakakibara2019}, which decreases its value if an accepting state is being approached.

\begin{defn}
	\label{def:rank_func}
	Let $ P=((X_P,\Sigma_P,\delta_P,x_{P,0}),F_P) $ be a product automaton. 
	A function $ \xi: X_P \to \mathbb{N} $ is a \textit{ranking function} for $ P $ if 
	\begin{align*}
	\forall x \in X_P,& \forall \sigma \in \Sigma_{P,uc}(x),\\
	&\xi(x) \geq \min \big\{ \xi(\delta_P(x,\sigma)) + \bar{I}_{F_P}(x), \alpha \big\}, 
	\end{align*}
	where $ \alpha > |X_P|-|F_P| $ and $ \bar{I}_{F_P}: X_P \to \{0,1\} $ is an indicator function such that $ \bar{I}_{F_P}(x) = 1 $ if and only if $ x \notin F_P $.  
\end{defn}


In \cite{Sakakibara2019}, 
we propose an algorithm to compute a ranking function for the product automaton.
Here, we show important results related to the ranking function $ \xi $ with the upper bound $ \alpha = |X_P| - |F_P| + 1 $.

\begin{prop}
	For any $ x \in X_P $, $ x \in F_P $ if and only if $ \xi(x) = 0 $. 
\end{prop}


\begin{prop}
	\label{prop:xi_valid}
	For any $ x \in X_P $, 
	\[ \xi(x) < \alpha \implies \exists s \in \Sigma_P^*, \delta_P(x,s) \in F_P. \]
\end{prop}

\begin{prop}
	\label{prop:suc_lower_rank}
	For any $ x \in X_P $, 
	\begin{align*}
	0 < \xi(x) & <\alpha\\
	\implies&\{ \sigma \in \Sigma_P(x): \xi(x) > \xi(\delta_P(x,\sigma)) \} \neq \emptyset.
	\end{align*}
\end{prop}


\begin{prop}
	\label{prop:uc_included}
	For any $ x \in X_P \setminus F_P $, 
	\[ \Sigma_{P,uc}(x) \subseteq \{ \sigma \in \Sigma_P(x): \xi(x) > \xi(\delta_P(x,\sigma)) \}. \]
\end{prop}
Proposition \ref{prop:suc_lower_rank} ensures that a lower-ranked successor always exists. 
Moreover, as mentioned in Proposition \ref{prop:uc_included}, each successor associated with an uncontrollable event has a lower rank than that of the current state.

In general, the existence of supervisors under partial observation depends on the controllability and observability of the specification language. 
Although these properties are defined by means of languages in the conventional supervisory control theory \citep{Cassandras2008}, here we characterize them with the ranking function.
\begin{defn}
	The product automaton $ P $ is said to be \textit{controllable} (with respect to $ \xi $) if $ \xi(x_{P,0}) < \alpha $. 
\end{defn}	

\begin{defn}
	The product automaton $ P $ is said to be \textit{observable} (with respect to $ \xi $ and $ \mathcal{P} $) if 
	\begin{align*}
		\forall s, s' \in \mathcal{L}(P), \  \forall & \sigma \in \Sigma_P, \\
		\mathcal{P}(s) = \mathcal{P}(s') &\land \xi(\delta_P(x_{P,0}, s)) > \xi(\delta_P(x_{P,0},s\sigma)) \\
		&\land s'\sigma \in \mathcal{L}(P) \\
		\implies & \xi(\delta_P(x_{P,0}, s')) > \xi(\delta_P(x_{P,0},s'\sigma)). 
	\end{align*}
\end{defn}
If the context is clear, we just say $ P $ is controllable or observable without referring to $ \xi $ and $ \mathcal{P} $.
The observability condition requires that, if an event is defined after different sequences with the same observation, then all of the transitions triggered by the event agree with each other in a sense of whether the product automaton gets closer to accepting states or not. 

We characterize transitions of the product automaton with respect to the ranking function. 
Let $ (x,\sigma,x') \in X_P \times \Sigma_P \times X_P $ be a transition defined in the product automaton.
\begin{itemize}
	\item $ (x,\sigma,x') $ is \textit{legal} (with respect to $ \xi $) if $ \xi(x) > \xi(x') $. 
	\item $ (x,\sigma,x') $ is \textit{neutral} (with respect to $ \xi $) if $ \xi(x) \leq \xi(x') < \alpha $. 
	\item $ (x,\sigma,x') $ is \textit{illegal} (with respect to $ \xi $) if $ \xi(x') = \alpha $. 
\end{itemize}

It is possible to lead the product automaton to reach an accepting state 
if we always choose legal transitions. 
On the other hand, however, 
we are likely to obtain more \textit{permissive} supervisors if we allow not only legal transitions but also neutral ones to be enabled. 
Permissiveness is one of the most important concepts in the supervisory control theory, where 
we often aim to design a supervisor that enables as many events as possible.

\section{ON-LINE SUPERVISORY CONTROL UNDER PARTIAL OBSERVATION}
\label{sec:on-line}
In this section, we explain the on-line supervisory control scheme for Problem \ref{prob:main}, given the product automaton $ P $ and the ranking function $ \xi $. 
Notice that we have a tradeoff between permissiveness of the supervisor and achievement of the specification. 
The supervisor becomes more permissive if it enables events triggering neutral transitions. 
However, infinitely many occurrences of neutral transitions results in livelock, i.e., the product automaton may stay within states $ x $ with $ 0 < \xi(x) < \alpha  $ while it always holds the possibility of reaching an accepting state but actually suspends going there.


To take the tradeoff into consideration, we introduce a criterion for how many neutral transitions we allow to be enabled.
More precisely, the supervisor we design determines its control action on-line, being aware of 
a time-varying permissiveness level, 
which is referred to together with the ranking function to improve permissiveness of supervisors. 
Here, we introduce a function that quantifies a permissiveness level. 

\begin{defn}
	\label{def:energy}
	A \textit{permissiveness function} is a function $ \eta:\mathbb{N} \to \mathbb{R} $ that satisfies the following three conditions. 
	\begin{enumerate}
		\item $ \eta(0) \leq \alpha $;
		\item $ \eta(k) \geq \eta(k+1) $ for any $ k \in \mathbb{N} $;
		\item $ \eta(\bar{k}) = 0 $ for some $ \bar{k} \in \mathbb{N} $. 
	\end{enumerate}
\end{defn}
That is, the permissiveness level decreases as time goes by and will eventually be exhausted.

With partially observed information, 
the supervisor cannot know which state the product automaton is currently in. 
We define the unobservable reach \citep{Hadj-Alouane1996} from a state $ x \in X_P $ under an event subset $ \Sigma' \subseteq \Sigma_P $ as follows. 
\begin{align*}
\mathsf{UR}_{\Sigma'}(x) =& \{ x' \in X_P: u \in (\Sigma_{uo} \cap \Sigma')^*, x' = \delta_P(x,u) \}. 
\end{align*}
For a subset $ X' \subseteq X_P $, let $ \mathsf{UR}_{\Sigma'}(X') = \bigcup_{x \in X'}\mathsf{UR}_{\Sigma'}(x) $. 
The set of \textit{next states} after each observation is defined recursively as follows: for $ s_o \in \mathcal{P}(\mathcal{L}(P)) $ and $ \sigma_o \in \Sigma_o $,
\begin{align*}
\mathsf{NS}(\varepsilon) &= \{ x_{P,0} \}, \\
\mathsf{NS}(s_o \sigma_o) &= \{ x' = \delta_P(x,\sigma_o) \in X_P: x \in \mathsf{UR}_{\Sigma_P}(\mathsf{NS}(s_o)) \}.
\end{align*}

The key idea of our on-line control scheme is made up of the following two rules.
First, we always enable events that trigger legal transitions no matter which state the DES is in or how much the permissiveness level currently is. 
Second, it is possible to allow neutral transitions if we have enough permissiveness level. 
These concepts are realized by two different control actions given by, for each $ x \in X_P $ and $ k \in \mathbb{N} $, 
\begin{align*}
\hat{\gamma}_{leg}(x)
=&\{ \sigma \in \Sigma_P: \exists s \in \Sigma_{uo}^*, \delta_P(x,s)! \land \delta_P(x,s\sigma)! \\ 
& \qquad \qquad \qquad \quad \land \xi(\delta_P(x,s)) > \xi(\delta_P(x,s\sigma)) \}, \\
\hat{\gamma}_{per}(x,k) =& \{ \sigma \in \Sigma_P: \forall s \in \Sigma_{uo}^*, \\ 
&\qquad\qquad \delta_P(x,s\sigma)! \implies \xi (\delta_P(x,s\sigma)) < \eta(k) \}. 
\end{align*}
Applying $ \hat{\gamma}_{leg}(x) $ results in occurrences of only legal transitions within $ \mathsf{UR}_{\Sigma_P} (x) $  
while the supervisor enables events in $ \hat{\gamma}_{per}(x) $, 
which may trigger neutral transitions, if the permissiveness level is high. 
The supervisor refers to the permissiveness function to see whether such additional events are acceptable currently or not.
From Proposition \ref{prop:suc_lower_rank}, we have the following proposition. 

\begin{prop}
	\label{prop:leg_nonempty}
	If the product automaton is controllable, then for any $ x \in X_P \setminus F_P $, we have $ \hat{\gamma}_{leg}(x) \neq \emptyset $. 
\end{prop}

\begin{prop}
	Assume that the product automaton is controllable and observable. 
	Then, for any $ x \in X_P \setminus F_P $, we have 
	\begin{multline*}
	\forall \sigma \in \hat{\gamma}_{leg}(x), \forall s \in \Sigma_{uo}^*, \\
	\qquad\delta_P(x,s)! \land \delta_P(x,s\sigma)! \implies \xi(\delta_P(x,s)) > \xi(\delta_P(x,s\sigma)).
	\end{multline*}
\end{prop}

\begin{pf}
	Let $ x \in X_P \setminus F_P $.
	From Proposition \ref{prop:leg_nonempty}, then, $ \hat{\gamma}_{leg}(x) \neq \emptyset $. 
	Suppose that, for some $ \sigma \in \hat{\gamma}_{leg}(x) $, we have 
	\begin{equation}
	\label{eq:leg_suppose}
	\begin{aligned}
	\exists s \in \Sigma_{uo}^*, &\delta_P(x,s)! \land \delta_P(x,s\sigma)! \\
	&\land \xi(\delta_P(x,s)) \leq \xi(\delta_P(x,s\sigma)).
	\end{aligned}
	\end{equation}
	From the definition of $ \hat{\gamma}_{leg}(x) $, the event $ \sigma $ satisfies 
	\begin{equation}
	\label{eq:leg_definition}
	\begin{aligned}
	\exists s' \in \Sigma_{uo}^*, &\delta_P(x,s')! \land \delta_P(x,s'\sigma)! \\
	&\land \xi(\delta_P(x,s)) > \xi(\delta_P(x,s\sigma)).
	\end{aligned}
	\end{equation}
	Let $ t_x \in \mathcal{L}(P) $ such that $ \delta_P(x_{P,0},t_x) $. 
	Then, we have $ \mathcal{P}(\mathcal{L}(t_xs)) = \mathcal{P}(\mathcal{L}(t_xs')) $ but for the event $ \sigma $ both Eqs.~\eqref{eq:leg_suppose} and \eqref{eq:leg_definition} hold, which contradicts the assumption that the product automaton is observable. 
\end{pf}

In the on-line control scheme, the supervisor keeps the set $ \mathsf{NS}(s_o) $ of states where the product automaton is estimated to be from the observation $ s_o \in \mathcal{P}(\mathcal{L}(P)) $. 
After observing an observable event $ \sigma_o \in \Sigma_o $, the supervisor updates the set to $ \mathsf{NS}(s_o \sigma_o) $ and computes a control pattern to be applied for the control of $ \mathsf{UR}_\Sigma(\mathsf{NS}(s_o\sigma_o)) $. 
Then, based on the idea mentioned above, the supervisor $ \hat{\mathcal{S}} $ computes a control pattern satisfying, for each $ s_o \in \mathcal{P}(\mathcal{L}(P)) $, 
\begin{equation}
\label{eq:olsup}
\hat{\mathcal{S}}(s_o) = \hat{\gamma}_{leg}(\mathsf{NS}(s_o)) \cup \hat{\gamma}_{per}(\mathsf{NS}(s_o),\|s_o\|),
\end{equation}
where 
\begin{align*}
\hat{\gamma}_{leg}(\mathsf{NS}(s_o)) = \bigcup_{x \in \mathsf{NS}(s_o)} \hat{\gamma}_{leg}(x),  \\
\hat{\gamma}_{per}(\mathsf{NS}(s_o)) = \bigcap_{x \in \mathsf{NS}(s_o)} \hat{\gamma}_{per}(x). 
\end{align*}

\begin{small}

\begin{algorithm}[!tb]
	\centering
	\caption{An on-line supervisory control algorithm for a DES $ G $ under an scLTL constraint $ \varphi $, given the product automaton $ P=((X_P,\Sigma_P,\delta_P,x_{P,0}),$ $F_P) $, a ranking function $ \xi:X_P \to \mathbb{N} $, and a permissiveness function $ \eta $} \label{alg:VLP}
	\begin{algorithmic}[1]
		\If{$ P $ is controllable and observable} 
		\State $ NS \gets \{ x_{P,0} \} $, $ k \gets 0 $
		\While{$ NS \not \subseteq F_P $} 
			\State $ \gamma_k \gets \emptyset, \bar{\gamma}_k \gets \emptyset, UR \gets \emptyset, NS' \gets \emptyset $
			\ForAll{$ x \in NS $}
				\State \Call{LegExpand}{$ x, \gamma_k, \bar{\gamma}_k, UR, NS' $}
			\EndFor
			\ForAll{$ \sigma \in \Sigma_{P,c} \setminus (\gamma_k \cup \bar{\gamma}_k) $}
				\State $ UR_\sigma \gets \emptyset, NS_\sigma' \gets \emptyset, \mathtt{st} \gets \mathrm{false} $
				\ForAll{$ x \in NS $}
					\State \Call{ReExpand}{$ x, \gamma_k \cup \{\sigma\}, UR_\sigma, NS_\sigma', \eta(k), $ $\mathtt{st} $}
					\If{$ \mathtt{st} $}
						\State \textbf{break}
					\EndIf
				\EndFor
				\If{$ \neg\mathtt{st} $}
					\State $ \gamma_k \gets \gamma_k \cup \{\sigma\} $
					\State $ UR \gets UR \cup UR_\sigma $
					\State $ NS' \gets NS' \cup NS'_\sigma $
				\EndIf
			\EndFor
			\State $ \sigma_o \gets \mathsf{observe}(G,\gamma_k) $
			\State $ NS \gets \{ x' \in NS': x \in UR, x' =\delta_P(x,\sigma_o) \} $
			\State $ k \gets k+1 $
		\EndWhile \label{line:online_end}
		\Else 
		\State There is no supervisor for $ G $ that guarantees $ \varphi $ under partial observation. 
		\EndIf
	\end{algorithmic}
%
	\begin{algorithmic}[1]
		\Function{LegExpand}{$ x \in X_P, \gamma, \bar{\gamma}, UR, NS' $}
		\If{$ x \notin UR $}
		\State $ UR \gets UR \cup \{x\} $
		\ForAll{$ \sigma \in \Sigma_P(x) $}
		\If{$ \xi(x) > \xi(\delta_P(x,\sigma)) $}
		\State $ \gamma \gets \gamma \cup \{\sigma\} $
		\If{$ \sigma \in \Sigma_{uo} $}
		\State \Call{LegExpand}{$ \delta_P(x,\sigma), \gamma,\bar{\gamma}, UR, $ $NS' $}
		\Else
		\State $ NS' \gets NS' \cup \{\delta_P(x,\sigma)\} $
		\EndIf
		\ElsIf{$ \xi(\delta_P(x,\sigma))= \alpha $}
		\State $ \bar{\gamma} \gets \bar{\gamma} \cup \{\sigma\} $
		\EndIf
		\EndFor
		\EndIf
		\EndFunction
	\end{algorithmic}
	\begin{algorithmic}[1]
		\Function{ReExpand}{$ x, \gamma, UR, NS',\eta_k, \mathtt{st} $}
		\If{$ x \notin UR \land \neg \mathtt{st} $}
		\State $ UR \gets UR \cup \{x\} $
		\ForAll{$ \sigma \in \gamma \cap \Sigma_P(x) $}
		\If{$ \xi(\delta_P(x,\sigma)) < \eta_k $}
		\If{$ \sigma \in \Sigma_{uo} $}
		\State \Call{ReExpand}{$ \delta_P(x,\sigma),\gamma, UR, NS', \eta_k, \mathtt{st} $}
		\Else
		\State $ NS' \gets NS' \cup \{\delta_P(x,\sigma)\} $
		\EndIf
		\Else
		\State $ \mathtt{st} \gets \mathrm{true} $ 
		\EndIf
		\EndFor
		\EndIf
		\EndFunction
	\end{algorithmic}
\end{algorithm}
\end{small}
From Proposition \ref{prop:leg_nonempty}, we have the following proposition. 
\begin{prop}
	\label{prop:leg_NS_nonempty}
	If the product automaton is controllable, then for any $ s_o \in \mathcal{P}(\mathcal{L}(P)) $, $ \hat{\gamma}_{leg}(\mathsf{NS}(s_o)) \neq \emptyset $. 
\end{prop}

\begin{prop}
	\label{prop:URconsistency}
	Assume that the product automaton is controllable and observable. 
	Then, for any $ s_o \in \mathcal{P}(\mathcal{L}(P)) $ and any $ x \in \mathsf{NS}(s_o) $, 
	\begin{equation}
	\forall x' \in \mathsf{UR}_{\Sigma_P} (x), 
	\forall \sigma \in \hat{\gamma}_{leg}(\mathsf{NS}({s_o})) \setminus \hat{\gamma}_{leg}(x), 
	\neg \delta_P(x',\sigma)!.
	\end{equation}
\end{prop}

\begin{pf}
	We prove the proposition by contradiction. 
	Let $ s_o \in \mathcal{P}(\mathcal{L}(P)) $. 
	By the controllability of $ P $ and Proposition \ref{prop:leg_NS_nonempty}, we have $ \hat{\gamma}_{leg}(\mathsf{NS}(s_o)) \neq \emptyset $. 
	Suppose that, for some $ x' \in \mathsf{UR}_{\Sigma_P}(x) $, there exists $ \sigma \in \hat{\gamma}_{leg}(\mathsf{NS}(s_o)) \setminus \hat{\gamma}_{leg}(x) $ with $ \delta_P(x',\sigma)! $.
	This means that for some $ y \in \mathsf{NS}(s_o) $ with $ y \neq x $, we have $ \sigma \in \hat{\gamma}_{leg}(y) $ and $ \delta_P(x', \sigma)! $. 
	Equivalently, both of the following conditions hold for some $ \sigma $:
	\begin{align*}
	1)\ &\exists t_x \in \mathcal{P}^{-1}(s_o), x = \delta_P(x_{P,0},t_x) \land \exists u \in \Sigma_{uo}^*, \delta_P(x,u\sigma)!.\\
	2) \ &\exists t_y \in \mathcal{P}^{-1}(s_o), y = \delta_P(x_{P,0},t_y) \land \exists s_y \in \Sigma_{uo}^*, \delta_P(y,s_y)! \\
	&\land \delta_P(y,s_y\sigma)! \land \xi(\delta_P(y,s_y)) > \xi (\delta_P(y,{s_y \sigma})). 
	\end{align*}
	In the above conditions, we have $ \mathcal{P}(t_xu) = \mathcal{P}(t_x) \varepsilon = s_o $ and $ \mathcal{P}(t_yy_s) = \mathcal{P}(t_y) \varepsilon = s_o $.
	To sum up, we have 
	\begin{align*}
	\exists t_x', t_y' & \in \mathcal{L}(P),\ \mathcal{P} (t_x') = \mathcal{P}(t_y') \land t_y'\sigma \in \mathcal{L}(P)  \\ &\land \xi (\delta_P(y,{t_y'})) > \xi(\delta_P(y,{t_y'\sigma})) \land t_x' \sigma \in \mathcal{L}(P).
	\end{align*}
	On the other hand, however, it holds that $ \xi(\delta_P(x,{t_x'})) \leq \xi(\delta_P(x,{t_x'\sigma})) $ because $ \sigma \notin \hat{\gamma}_{leg}(x) $.
	We now have the contradiction to the observability condition of the product automaton.
	%
\end{pf}
From Proposition \ref{prop:URconsistency}, we have the following lemma.
\begin{lem}
	If the product automaton is controllable and observable, we have, 
	for any $ s_o \in \mathcal{P}(\mathcal{L}(P)) $,  
	\[ \forall x \in \mathsf{NS}({s_o}), \ \mathsf{UR}_{\hat{\gamma}_{leg}(\mathsf{NS}({s_o}))}(x) = \mathsf{UR}_{\hat{\gamma}_{leg}(x)}(x). \]
\end{lem}

\begin{cor}
	If the product automaton is observable, 
	then for any $ s_o \in \mathcal{P}(\mathcal{L}(P)) $,  
	\[ \forall x \in \mathsf{NS}({s_o}), \ \max_{ x' \in \mathsf{UR}_{\hat{\gamma}_{leg}(\mathsf{NS}({s_o}))}(x)} \xi(x') < \xi(x). \]
\end{cor}

\subsection{On-line Control Algorithm}

\subsubsection{Main part.}
The on-line supervisory control scheme is described in Algorithm \ref{alg:VLP}.
The supervisor keeps the sets $ NS $ of next states after the observation so far, which is initialized with $ \{ x_{P,0} \} $. 
At each step $ k $, the on-line supervisor computes a control pattern $ \gamma_k $ that will be applied to $ NS $, the corresponding unobservable reach $ UR $, and the set $ NS' $ of \textit{new} next states after potential occurrences of observable events. 
The functions \textproc{LegExpand} and \textproc{ReExpand} compute $ \hat{\gamma}_{leg}(NS) $ and $ \hat{\gamma}_{per}(NS) $, respectively (but for now we skip the detailed explanations).
After computed, the control pattern $ \gamma_k $ is issued to the DES $ G $, which executes one of the enabled events.
The new observation is represented by the function $ \mathsf{observe}(G, \gamma_k) $, according to which  
the supervisor updates information related to the memory $ NS $ and time step $ k $ and goes on to determine the next control action. 


\subsubsection{Subfunctions.}
In the main part of Algorithm \ref{alg:VLP}, we first compute $ \hat{\gamma}_{leg}(NS) $ by the function \textproc{LegExpand}. 
Then, by the function \textproc{ReExpand}, we additionally examine if other controllable events can be added to the next control pattern. 
The functions \textproc{LegExpand} and \textproc{ReExpand} expands states in $ NS $ in a depth-first-search mannar until an observable event is detected.

The function \textproc{LegExpand} expands an input state $ x \in X_P $ and updates $ \gamma, \bar{\gamma}, UR $, and $ NS' $ if necessary.
When a legal transition with an event $ \sigma $ is detected during the search, the event is added to $ \gamma $.
After the call of \textproc{LegExpand} in Algorithm \ref{alg:VLP}, we have $ \hat{\gamma}_{leg}(NS) $ and the unobservable reach of $ NS $ under the control pattern $ \hat{\gamma}_{leg}(NS) $. 
We move on to the other function \textproc{ReExpand}, 
which examines each controllable event $ \sigma_c $ that has not been in $ \hat{\gamma}_{leg}(NS) $.
Unlike \textproc{LegExpand}, we have a global boolean variable $ \mathtt{st} $, initialized with false. 
It is necessary to examine all states that may be visited if $ \sigma_c $, the currently examined controllable event, is added to $ \gamma_k $.
If we find the event $ \sigma_c $ cannot be in $ \hat{\gamma}_{per}(NS) $, then the variable $ \mathtt{st} $ turns to be true.

\subsection{Correctness of Algorithm \ref{alg:VLP}}
The on-line supervisor refers to the current rank $ \xi(x) $ and the current permissiveness level $ \eta(k) $ to take into consideration the tradeoff between permissiveness and acceptance of the specification. 
More precisely, the more permissiveness level we have, the more neutral transitions we allow to be enabled. 
Since the permissiveness level decreases with the elapse of time, the supervisor enables less events as time goes by. 
In the following, we show the correctness of Algorithm \ref{alg:VLP}.

\begin{lem}
	If the product automaton is controllable, then 
	for any $ s_o \in \mathcal{P}(\mathcal{L}(P)) $, 
	$ \hat{\mathcal{S}}(s_o) \neq \emptyset $. 
\end{lem}

\begin{pf}
	By Eq.~(\ref{eq:olsup}), for any $ s_o \in \mathcal{P}(\mathcal{L}(P)) $, we have $ \hat{\mathcal{S}}(s_o) \supseteq \hat{\gamma}_{leg}(\mathsf{NS}(s_o)) $, which is always nonempty as mentioned in Proposition \ref{prop:leg_NS_nonempty}.
\end{pf}

\begin{lem}
	For any $ s_o \in \mathcal{P}(\mathcal{L}(P)) $ and any $ x \in \mathsf{UR}_\Sigma(\mathsf{NS}(s_o)) $,  
	we have $ \hat{\mathcal{S}}(s_o) \in \Gamma(J_G(x)) $.
\end{lem}

\begin{pf}
	From Proposition \ref{prop:uc_included}, for any $ x \in X_P $, $ \hat{\gamma}_{leg}(x) \in \Gamma(J_G(x)) $.
	Then, by Eq.~\eqref{eq:olsup} and the definition of $ \hat{\gamma}_{leg}(\mathsf{NS}(s_o)) $, the lemma holds.
\end{pf}

\begin{prop}
	\label{prop:eventually_empty}
	Assume that the product automaton $ P $ is observable. 
	For any $ x \in X_P $ with $ 0 < \xi(x) < \alpha $, there exists $ k \in \mathbb{N} $ such that, for any $ l \geq k $,  
	$ \hat{\gamma}_{per}(x,l) = \emptyset. $
\end{prop}

\begin{pf}
	Note that, for each $ x \in X_P $ and $ k \in \mathbb{N} $,
	\[ \hat{\gamma}_{per}(x,k) = \bigcap_{x' \in \mathsf{UR}_\Sigma(x)} \{ \sigma \in \Sigma_P(x'): \xi(\delta_P(x',\sigma)) < \eta(k) \}. \]
	By the second and third conditions of Definition \ref{def:energy}, on the other hand, there exists $ \bar{k} \in \mathbb{N} $ such that, for all $ \bar{l} \geq \bar{k} $, $ \eta(\bar{l})=0 $.	
	Since the ranking function $ \xi $ returns nonnegative values, $ \{ \sigma \in \Sigma_P: \xi(\delta_P(x,\sigma)) < 0 \} = \emptyset $ for any $ x \in X_P $.
	That is, we have $ \hat{\gamma}_{per}(x,\bar{l}) = \bigcap_{x' \in \mathsf{UR}_\Sigma(x)} \emptyset = \emptyset $ for all $ \bar{l} \geq \bar{k} $.
\end{pf}

\begin{lem}
	\label{lem:neutral_limited}
	Assume that the product automaton $ P $ is observable. 
	In the while loop of Algorithm \ref{alg:VLP}, 
	neutral transitions of the product automaton occur only finitely often.
\end{lem}

\begin{pf}
	We prove the lemma by contradiction. 
	Suppose that neutral transitions occur infinitely often.
	That is, there exists $ s \in \Sigma^\omega $ such that 
	\begin{equation}
	\label{eq:ass}
	\begin{array}{l}
	\!\!\!\!\forall k \in  \mathbb{N}, \exists l \geq k, \\
	\xi(\delta_P(x_{P,0}, s[0\ldots l])) \leq \xi( \delta_P(x_{P,0}, s[0\ldots l+1]) ).
	\end{array}
	\end{equation}
	Let $ \sigma_{l+1}\!:=\!s[l\!+\!1] $ for some $ l\! \in\! \mathbb{N} $ that satisfies the above inequality, $ x_{l+1}:= \delta_P(x_{P,0},s[0\ldots l]) $, and $ t_l := \mathcal{P}(s[0\ldots l]) $. 
	Then, we have 
	\[ \sigma_{l+1} \in \hat{\mathcal{S}}(t_l) = \hat{\gamma}_{leg}(\mathsf{NS}(t_l)) \cup \hat{\gamma}_{per}(\mathsf{NS}(t_l),\|t_l\|). \]
	Since transition $ (x_l, \sigma_{l+1}, x_{l+1}) $ is neutral, it holds that 
	\begin{equation}
	\label{eq:sigma_l_per}
	\sigma_{l+1} \in \hat{\gamma}_{per}(\mathsf{NS}(t_l),\|t_l\|).
	\end{equation}
	By Eq.~(\ref{eq:ass}), therefore, there exist infinitely many $ l \in \mathbb{N} $ satisfying Eq.~(\ref{eq:sigma_l_per}), which contradicts Proposition \ref{prop:eventually_empty}.
\end{pf}

\begin{lem}
	\label{lem:online_accepting}
	Assume that $ P $ is controllable and observable. 
	In Algorithm \ref{alg:VLP}, 
	an accepting state of the product automaton is eventually reached 
	under the control by the on-line supervisor $ \hat{\mathcal{S}} $.
\end{lem}

\begin{pf}
	Recall that $ \eta(0) \leq \alpha $ and that $ e $ is nonincreasing, as mentioned in Definition \ref{def:energy}.
	By the controllability of $ P $, we have $ \xi(x_{P,0}) < \alpha $. 
	Since the on-line supervisor $ \hat{\mathcal{S}} $ never allows illegal transitions, 
	we have, for any $ s \in \mathcal{L}(\hat{\mathcal{S}}/G) $, $ \xi(\delta_P(x_{P,0},s)) < \alpha $. 
	By Proposition \ref{prop:xi_valid}, then, it is always possible to lead the product automaton to an accepting state by some appropriate event sequence.
	From Lemma \ref{lem:neutral_limited}, while the on-line computation is running, the supervisor $ \hat{\mathcal{S}} $ observes neutral transitions only finitely often. 
	Let $ \bar{k} \in \mathbb{N} $ be the step index such that $ \eta(\bar{k}) = 0 $. 
	Then, for any observation $ s_o \in \mathcal{P}(\mathcal{L}(P)) $ with $ \|s_o\|>\bar{k} $, we have 
	$ \hat{\mathcal{S}}(s_o) = \hat{\gamma}_{leg}(\mathsf{NS}(s_o)) $.
	In other words, the supervisor $ \hat{\mathcal{S}} $ chooses only legal transitions after time step $ \bar{k} $. 
	Since the rank always decreases during each legal transition, eventually a state ranked as $ 0 $, namely, an accepting state is reached. 
\end{pf}

\begin{thm}
	$ \hat{\mathcal{S}}/G \models \varphi $ if $ P $ is controllable and observable. 
\end{thm}

\begin{pf}
	From Lemma \ref{lem:online_accepting}, the on-line supervisor forces the plant DES $ G $ to generate event sequences with which the product automaton eventually reaches an accepting state. 
	Note that, when an accepting state of the product automaton is reached, then the corresponding word is accepted by the DFA $ A_\varphi $. 
	By Eq.~(\ref{eq:acc_scLTL}), any run that has the corresponding history as a prefix satisfies the scLTL formula $ \varphi $.
\end{pf}

%




\section{Illustrative Example}
\label{sec:example}
\begin{figure}[!tb]
	\centering
	\includegraphics[bb=0 0 164 158,scale=0.56]{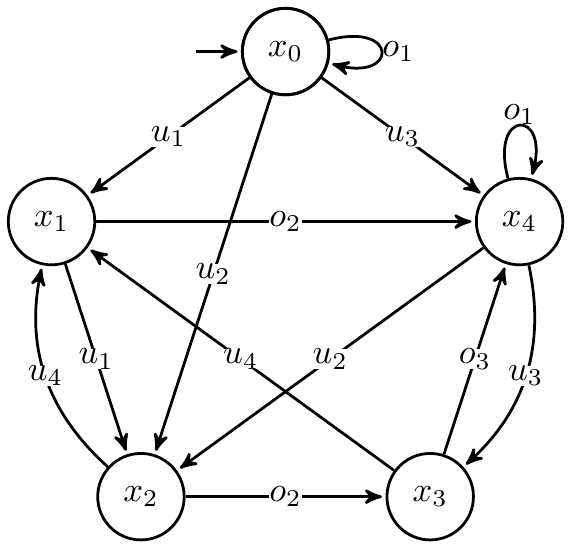} \hspace{10pt}
	\includegraphics[bb=0 0 218 188,scale=0.56]{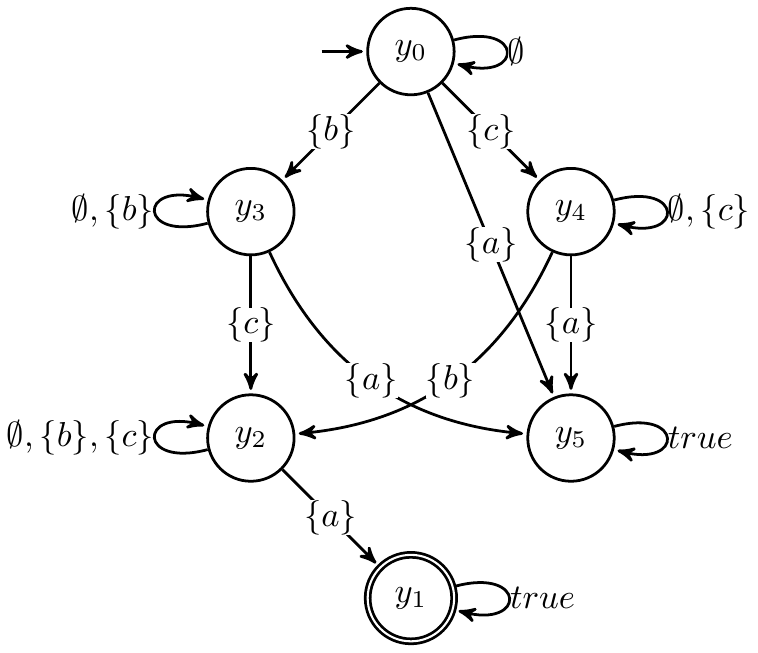}
	\caption{(Left) The DES $ G_{ex} $ discussed in Section \ref{sec:example}, where $ \Sigma_o = \{o_1,o_2,o_3\} $ and $ \Sigma_{uo}=\{u_1,u_2,u_3,u_4\} $. (Right) The DFA $ A_{\varphi_{ex}} $ translated from $ \varphi_{ex} $, where $ F_A = \{y_1\} $.}
	\label{fig:example}
\end{figure}

In this section, we demonstrate the proposed method with a simple example. 
Consider a DES $ G_{ex} $ depicted in the left of Fig.~\ref{fig:example}, where $ X = \{x_0, x_1, x_2, x_3, x_4 \} $; $ \Sigma_o = \{o_1,o_2,o_3\} $ and $ \Sigma_{uo}=\{u_1,u_2,u_3,u_4\} $; $ AP = \{a,b,c\} $; $ L(x_0)=L(x_4)=\emptyset $, $ L(x_1)=\{a\} $, $ L(x_2)=\{b\} $, and $ L(x_3) = \{c\} $. 
We assume that all events are controllable and the initial state is $ x_0 $.
For the DES $ G_{ex} $, we impose an scLTL formula 
$ \varphi_{ex} = \Diamond a \land \neg a \mathbf{U} b \land \neg a \mathbf{U} c $ 
as a control specification. 
$ \varphi_{ex} $ requires to eventually go to a state labeled with $ a $ after visiting both $ b $-states and $ c $-states.
We use a tool Spot\footnote{https://spot.lrde.epita.fr/} to translate $ \varphi_{ex} $ into a DFA, which is shown in the right of Fig.~\ref{fig:example}. 

\begin{figure}[!tb]
	\centering
	\includegraphics[bb=0 0 265 176, scale=0.65]{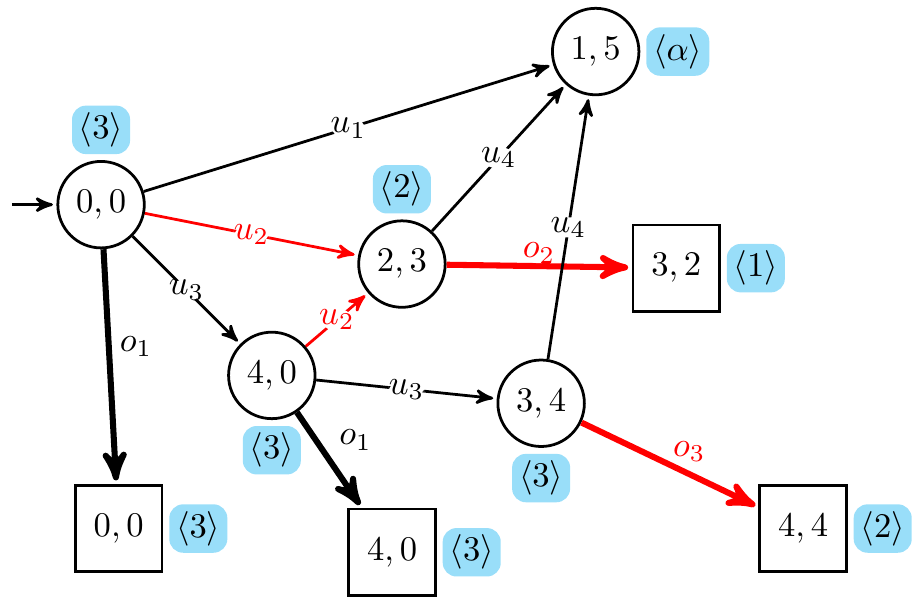}
	\caption{
		The computation tree of the unobservable reach at the initial step ($ k=0 $), 
		where circle and rectangle nodes represent states in $ UR $ and $ NS' $, respectively, 
		bold arrows are transitions triggered by an observable event, and 
		legal transitions are colored by red. 
		A node label of the form $ (i,j) $ stands for state $ (x_i,y_j) \in X_P $.
		The rank of each state is shown by an additional label of the form $ \langle \cdot \rangle $.
	} 
	\label{fig:URcomputation}
\end{figure}

Let $ \eta(k) = \max \{ -k+5, 0 \} $. 
Then, at the initial step of the on-line control scheme, the unobservable reach from $ NS = \{x_{P,0} = (x_0,y_0)\} $ is computed as shown in Fig.~\ref{fig:URcomputation}.
After the call of \textproc{LegExpand}, we have $ UR =\{ (x_0,y_0), (x_2,y_3) \} $, $ NS' = \{ (x_3,y_2) \} $, and $ \gamma_0 = \hat{\gamma}_{leg}((x_0,y_0)) = \{u_2,o_2\} $. 
Since the current permissiveness level is $ \eta(0) = 5 $, the function \textproc{ReExpand} adds $ u_3, o_1, o_3 $ to $ \gamma_0 $. 
The sets $ UR $ and $ NS' $ are also updated with the related states. 
We do not expand the state $ (x_1,y_5) $ any more because its rank hits the upper bound.

Assume that $ \mathsf{observe}(G_{ex},\gamma_0) = o_1 $, according to which the supervisor updates $ NS $ to $ \{(x_0,y_0), (x_4,y_0)\} $ and ends up obtaining $ \gamma_1 = \gamma_0 $ after the computation at the next step.
Then, assume that the event $ o_1 $ is observed again, i.e., $ \mathsf{observe}(G_{ex},\gamma_1) = o_1 $.
Although we start from the same set $ NS = \{(x_0,y_0), (x_4,y_0)\} $ as the previous step, we have a different result.
Since $ \eta(2) = 3 $, the function \textproc{ReExpand} does not add to the current control pattern $ \gamma_2 $ events triggering neutral transitions to states ranked as $ 3 $. 
At the end of the computation for $ k=2 $, we have $ \gamma_2 = \{u_2,o_2\} $ and $ NS' = \{(x_3,y_2)\} $. 

Similarly, after observing $ o_2 = \mathsf{observe}(G_{ex},\gamma_2) $, only legal transitions are enabled because all states reachable from $ (x_3,y_2) $ have a rank lower than $ \eta(3) = 2 $. 
At the end of the computation at step $ 3 $, we have $ NS' \subseteq F_P $.
No matter which event is observed, then, the supervisor stops the control.

\section{Conclusion}
\label{sec:conclusion}
We propose a novel on-line supervisory control scheme of partially observed DESs to achieve a control specification given by scLTL formulas. 
We introduce the controllability and observability based on the ranking function, which derives a sufficient condition for the existence of the on-line supervisor. 
In the on-line computation, the supervisor computes the unobservable reach after each observation and refers to the ranking function together with the permissiveness function.  
Depending on the permissiveness level at each step, the supervisor improves its permissiveness if possible. 
It is future work to extend the proposed scheme to cases of general or quantitative LTL and to establish a verification method of the observability.


\bibliography{bibdata}             

\begin{thebibliography}{15}
\providecommand{\natexlab}[1]{#1}
\providecommand{\url}[1]{\texttt{#1}}
\providecommand{\urlprefix}{URL }
\expandafter\ifx\csname urlstyle\endcsname\relax
  \providecommand{\doi}[1]{doi:\discretionary{}{}{}#1}\else
  \providecommand{\doi}{doi:\discretionary{}{}{}\begingroup
  \urlstyle{rm}\Url}\fi

\bibitem[{Baier and Katoen(2008)}]{Baier2008}
Baier, C. and Katoen, J.P. (2008).
\newblock \emph{{Principles of Model Checking}}.
\newblock MIT Press.

\bibitem[{Belta et~al.(2017)Belta, Yordanov, and {Aydin Gol}}]{Belta2017}
Belta, C., Yordanov, B., and {Aydin Gol}, E. (2017).
\newblock \emph{{Formal Methods for Discrete-Time Dynamical Systems}}.
\newblock Springer International Publishing.

\bibitem[{Cassandras and Lafortune(2008)}]{Cassandras2008}
Cassandras, G.C. and Lafortune, S. (2008).
\newblock \emph{{Introduction to Discrete Event Systems}}.
\newblock Springer US, 2 edition.

\bibitem[{Chung et~al.(1992)Chung, Lafortune, and Lin}]{Chung1992}
Chung, S.L., Lafortune, S., and Lin, F. (1992).
\newblock {Limited lookahead policies in supervisory control of discrete event
  systems}.
\newblock \emph{IEEE Trans.~Autom.~Control}, 37(12), 1921--1935.

\bibitem[{Chung et~al.(1993)Chung, Lafortune, and Lin}]{Chung1993}
Chung, S.L., Lafortune, S., and Lin, F. (1993).
\newblock {Recursive computation of limited lookahead supervisory controls for
  discrete event systems}.
\newblock \emph{Discret. Event Dyn. Syst. Theory Appl.}, 3(1), 71--100.

\bibitem[{Chung et~al.(1994)Chung, Lafortune, and Lin}]{Chung1994}
Chung, S.L., Lafortune, S., and Lin, F. (1994).
\newblock {Supervisory control using variable lookahead policies}.
\newblock \emph{Discret.~Event Dyn.~Syst.~Theory Appl.}, 4(3), 237--268.

\bibitem[{Grigorov and Rudie(2006)}]{Grigorov2006}
Grigorov, L. and Rudie, K. (2006).
\newblock {Near-optimal online control of dynamic discrete-event systems}.
\newblock \emph{Discret.~Event Dyn.~Syst.~Theory Appl.}, 16(4), 419--449.

\bibitem[{Hadj-Alouane et~al.(1996)Hadj-Alouane, Lafortune, and
  Lin}]{Hadj-Alouane1996}
Hadj-Alouane, N.B., Lafortune, S., and Lin, F. (1996).
\newblock {Centralized and distributed algorithms for on-line synthesis of
  maximal control policies under partial observation}.
\newblock \emph{Discret.~Event Dyn.~Syst.~Theory Appl.}, 6(4), 379--427.

\bibitem[{Heymann and Lin(1994)}]{Heymann1994}
Heymann, M. and Lin, F. (1994).
\newblock {On-line control of partially observed discrete event systems}.
\newblock \emph{Discret.~Event Dyn.~Syst.~Theory Appl.}, 4(3), 221--236.

\bibitem[{Jiang and Kumar(2006)}]{Jiang2006}
Jiang, S. and Kumar, R. (2006).
\newblock {Supervisory control of discrete event systems with CTL* temporal
  logic specifications}.
\newblock \emph{SIAM J. Control Optim.}, 44(6), 2079--2103.

\bibitem[{Kupferman and {Y. Vardi}(2001)}]{Kupferman2001}
Kupferman, O. and {Y. Vardi}, M. (2001).
\newblock {Model checking of safety properties}.
\newblock \emph{Formal Methods in System Design}, 19(3), 291--314.

\bibitem[{Prosser et~al.(1998)Prosser, Kam, and Kwatny}]{Prosser1998}
Prosser, J.H., Kam, M., and Kwatny, H.G. (1998).
\newblock {Online supervisor synthesis for partially observed discrete-event
  systems}.
\newblock \emph{IEEE Trans.~Autom.~Control}, 43(11), 1630--1634.

\bibitem[{Ramadge and Wonham(1987)}]{Ramadge1987}
Ramadge, P.J. and Wonham, W.M. (1987).
\newblock {Supervisory control of a class of discrete event processes}.
\newblock \emph{SIAM J. Control Optim.}, 25(1), 475--498.

\bibitem[{Sakakibara and Ushio(2020)}]{Sakakibara2019}
Sakakibara, A. and Ushio, T. (2020).
\newblock {On-line permissive supervisory control of discrete event systems for
  scLTL specifications}.
\newblock \emph{IEEE Control Systems Letters}, 4(3), 530--535.

\bibitem[{Tumova and Dimarogonas(2016)}]{Tumova2016}
Tumova, J. and Dimarogonas, D.V. (2016).
\newblock {Multi-agent planning under local LTL specifications and event-based
  synchronization}.
\newblock \emph{Automatica}, 70, 239--248.

\end{thebibliography}
                                                   







\appendix
\renewcommand{\thealgorithm}{\Alph{section}.\arabic{algorithm}}

\section{Computation of Ranking Function}    
\label{app:offline}
Here, we briefly explain the results in \cite{Sakakibara2019}, where we propose an algorithm to compute a ranking function given the product automaton $ P $ of the DES and the DFA.
We obtain a ranking function by Algorithm \ref{alg:pm}.

\begin{algorithm}[htb]
	\centering
	\caption{Computation of a ranking function} \label{alg:pm}
	\begin{algorithmic}[1]
		\Require{
			A product automaton $ P = ((X_P,\Sigma_P,\delta_P,x_{P,0}),$ $F_P) $
		}
		\Ensure{A ranking function $ \xi: X_P \to \mathbb{N} $}
		\State $ \alpha \gets |X_P|-|F_P|+1 $ \label{line:pm_init_start}
		\ForAll{$ x \in X_P $} 
		\State $ \xi(x) \gets 0 $
		\EndFor \label{line:pm_init_end}
		\While{$ \exists x \in X_P $ s.t. $ \xi(x)  < \mathsf{up}_\alpha ( \hat{\xi}(x) ,x) $} \label{line:pm_update_start}
		\State $ \xi (x) \gets \mathsf{up}_\alpha (\hat{\xi}(x),x)$
		\EndWhile \label{line:pm_update_end}
	\end{algorithmic}
\end{algorithm}


As initialization, we set $ \xi(x) = 0 $ for each $ x \in X_P $ and $ \alpha = |X_P|-|F_P|+1 $. 
Then, we go on to update the values of $ \xi $ by using functions $ \hat{\xi}: X_P \to \mathbb{N} $ 
and $ \mathsf{up}_\alpha : \mathbb{N} \times X_P \to \mathbb{N} $, defined as follows.
For each $ x \in X_P $, 
\begin{align*}
&\hat{\xi}(x) =
\begin{cases}
\displaystyle
\min_{\sigma \in \Sigma_{P,c}} \xi( \delta_P(x,\sigma) ) & \text{if }\Sigma_{P,uc}(x) = \emptyset, \\
\displaystyle
\max_{\sigma \in \Sigma_{P,uc}} \xi(\delta_P(x,\sigma)) & \text{otherwise. } 
\end{cases}
\end{align*}
For any $ r \in \mathbb{N} $ and $ x \in X_P $, 
\begin{align*}
&\mathsf{up}_\alpha (r, x) 
= 
\begin{cases}
r+1 & \text{ if } x \notin F_P \land r < \alpha ,  \\
r & \text{ otherwise. }
\end{cases}
\end{align*}
To sum up, the current rank is incremented if the current state is not accepting with at least one uncontrollable event defined and the rank has not hit the upper bound;
otherwise the rank does not change.

\end{document}